\documentstyle[11pt]{article}

\input psfig

\renewcommand{\headheight}{20mm}
\renewcommand{\thefootnote}{\fnsymbol{footnote}}
\def\xslide#1#2#3#4#5#6{\centerline{\psfig
{figure=#1,height=#2,bbllx=#3bp,bblly=#4bp,bburx=#5bp,bbury=#6bp,clip=}}}

\begin{document}
\begin{large}

\long\def\equalign
#1\end#2{\def\\{\cr\noalign{\smallskip}}
    \refstepcounter{equation}$$ \vcenter{\tabskip=5pt
    \halign{\hfil$\displaystyle{##}$&&$\displaystyle{{}##}$\hfil\cr
    #1 \cr}}\end{#2}}
\let\endequalign\endequation \def\theequation{\arabic{equation}}

\def\gapp{\raisebox{-.5ex}{$\stackrel{>}{\scriptstyle\sim}$}}

\def\etal{{\it et al.}}
\def\ie{$ i. e. $}

\def\pamu{\partial^{\mu}}
\def\gmu{\gamma_{\mu}}

\def\psib{\overline{\psi}} \def\siga{\sigma_a} \def\g5{\gamma_5}
\def\s2{{G_S \over 2}} \def\v2{{G_V \over 2}} \def\gs{G_S}
\def\gv{G_V}

\def\sa{\sum_{a=0}^{3}}
\def\sp{\int {d^4 p \over (2\pi)^4}}
\def\sofp{S(p) }

\def\Tr{\hbox{Tr }}
\def\tr{\hbox{tr }}
\def\Sp{\hbox{Sp }}

\def\nc{N_c}
\def\nf{N_f}
\def\su2f{$\hbox{SU(2)}_{flavour}$}

\def\qb{\overline q}
\def\ppq{p + Q}
\def\pmq{p}
\def\qv{(\bf q \,)}
\def\aq{(q)}
\def\omp{\omega_{+}}
\def\omm{\omega_{-}}


\vspace{1cm}
\begin{center}

\begin{Large}
\bf{OSCILLATIONS OF THE STATIC MESON FIELDS AT FINITE BARYON DENSITY}
\end{Large}  \\[10mm]

\setcounter{footnote}{00}
\def\thefootnote{\arabic{footnote}}
Wojciech Florkowski$^{a}$ and Bengt Friman$^{b,c}$
\\[10mm]

\rm\begin{tabular}{ll}
$^a$ &H. Niewodnicza\'nski Institute of Nuclear Physics\\
&ul. Radzikowskiego 152, 31-342 Krak\'ow, Poland \\[3mm]
$^b$ &Gesellschaft f\"ur Schwerionenforschung GSI\\
&Postfach 110552, D-64220 Darmstadt, Germany\\[3mm]
$^c$ &Institut f\"ur Kernphysik, Technische Hochschule Darmstadt\\
& D-64289 Darmstadt, Germany\\[5mm]
\end{tabular}
\end{center}

\noindent{\bf Abstract:} The spatial dependence of static meson
correlation functions at finite baryon density is studied in the
Nambu -- Jona-Lasinio model. In contrast to the finite temperature
case, we find that the correlation functions at finite density are
not screened but exhibit long-range oscillations. The observed phenomenon is
analogous to the Friedel oscillations in a degenerate electron gas.

\setcounter{equation}{0}


\newpage
\renewcommand{\headheight}{-20mm}
\centerline{\bf 1. Introduction}
\bigskip

Recently, there is a growing interest in studying the properties of
hadronic correlation functions. The correlation functions form a
common ground, where QCD inspired effective theories can be tested
against each other as well as against lattice results, where available
\cite{ES}.

The hadronic correlation functions have been
intensively studied in lattice simulations of QCD at finite
temperatures \cite{DTK,SG,GRH,KB}. The results of these simulations
show that at large distances the correlation functions decrease
exponentially. The screening masses characterizing this type of
behaviour are, at $T > T_c$, degenerate for chiral partners. This
reflects the restoration of chiral symmetry at high temperatures.
Moreover, as it was pointed out by Eletskii and Ioffe \cite{EI}
(see also \cite{FF1}), the
values of the screening masses are close to $2\pi T$ for mesons and
$3\pi T$ for baryons. Such results are typical for a gas of
non-interacting quarks. Therefore, the lattice results show that at
high temperatures quarks behave, to large extent, as free particles.

The results of the lattice simulations have been interpreted in terms
of simple models by several authors:
Gocksch \cite{AG} argues that the hadronic screening masses can be well
described in the linear sigma model; Hansson and Zahed \cite{HZ}
claim that the static correlation functions in high-temperature QCD
can be well calculated from an equivalent problem of non-relativistic
quarks in a dimensionally reduced theory; similarly, V. Koch et al.
\cite{KSBJ} argue that the propagation of a light fermion in a spatial
direction at high temperatures can be described effectively by a
two-dimensional Schr\"odinger equation. 

In Ref. \cite{FF2} the temperature dependence of the meson screening
masses was computed and discussed in the  framework of the Nambu --
Jona-Lasinio (NJL) model. The results are in qualitative
agreement with the lattice simulations.  In the present paper, we
explore the static meson correlation  functions at finite baryon
density, using the methods developed in \cite{FF1,FF2}. In this way,
we complement and generalize our previous work. We note that, for
studying systems at finite baryon density one has to rely
on models, since lattice results are available only for vanishing
density.

The NJL model was introduced already in the early sixties as a theory
of interacting nucleons \cite{NJL}. Later it was reformulated in terms
of quark degrees of  freedom. Numerous calculations
demonstrate the success of the model in describing  hadronic data
(for a recent review see one of the articles in \cite{NJLR}). Two
general characteristics of QCD are described by the model,  namely
the chiral invariance which is spontaneously broken in the ground
state and the chiral phase transition (the chiral symmetry is restored
at high temperature or density). On the other hand, the model is not
renormalizable and not confining. Thus, it is an effective theory, 
valid only at low energies for phenomena, which are controlled by the 
chiral symmetry of QCD.

Since the NJL model exhibits a chiral phase transition, it is very
instructive to use it in order to study the temperature (density)
dependence of various physical quantities like, e.g., the quark
condensate \cite{HK85}, the pion decay constant \cite{BMZ87a,HK87},
and the dynamic masses of hadrons \cite{HK85,HK87,BMZ87b}. The
model has also been used to investigate the thermodynamic properties
of the quark-meson plasma near the chiral phase transition
\cite{HKZ}.

In this paper we study the in-medium static meson correlation
functions at finite baryon density. In contrast to the behaviour
known from the high temperature calculations, we find that the
correlation functions at finite baryon density are not screened
but oscillate in space. This behaviour is analogous to the Friedel
oscillations in a degenerate electron gas \cite{FW}; they are
caused by the existence of the sharp Fermi surface.

We find that the NJL approach to the correlation functions breaks 
down already at moderate densities. Thus, to complete the picture 
we compute the meson correlation function at higher densities in
perturbative QCD. Keeping only the leading term, we find oscillations 
with a period $\delta r = \pi/p_F$, where $p_F$ is the Fermi momentum 
of the quark sea.

The paper is organized as follows. In the next section we define the
model. In Section 3 the details of the calculations are presented.
Section 4 contains our results on the density dependence of the quark
and meson masses. In Section 5 we discuss our results on the
correlation function. The results based on the perturbative QCD are
presented in Section 6. We summarize the paper in Section 7.

\bigskip
\centerline{\bf 2. Definition of the model}
\bigskip

 Our calculations are based on the following form of the Lagrangian

\begin{equalign}
\label{lag}
{\cal L} & = &\psib (i \gmu \pamu - m) \psi +
\sa \s2 \left[(\psib \siga \psi)^2 +(\psib i\g5 \siga \psi)^2\right].
\end{equalign}

\noindent Here $\psi$ is the Dirac field with additional flavour
($\nf$=2) and colour ($\nc$=3) degrees of freedom, $\siga$ are the
Pauli matrices (with $\sigma_0$=1), $\gs$ is the coupling constant,
and $m$ is the current quark mass ($m_u = m_d = m$).

In the chiral limit, $m \rightarrow 0$, the Lagrangian (\ref{lag}) is
invariant under the unitary transformations $\hbox{U}_V$(1),
$\hbox{U}_A$(1) and the chiral symmetry
$\hbox{SU}_L$(2)$\times\hbox{SU}_R$(2). These are fundamental
symmetries of QCD, the underlying theory of strong interactions. In
the QCD vacuum the axial U$_A$(1) symmetry is broken due to the
instanton effects \cite{THOOFT} and 
$\hbox{SU}_L$(2)$\times\hbox{SU}_R$(2) is spontaneously broken down to
$\hbox{SU}_V$(2). In the Nambu--Jona-Lasinio model the spontaneous
breaking of the chiral symmetry is reproduced. The explicit
breaking of the $\hbox{U}_A$(1) symmetry, accounting for the instanton
effects, can be modelled by adding an extra term to the Lagrangian
(\ref{lag}). In order to keep things as simple as possible, we do not
take this term into account here. This does not affect our results,
since we do not consider the $\eta - \eta'$ channel. 

The self-energy of quarks is obtained in the Hartree-Fock
approximation by a self-consistent solution of the Schwinger-Dyson
equation

\begin{eqnarray}
\label{self1}
\Sigma & = & \gs \, i \sa \sp \left[
\siga \Tr[\siga \sofp] -
\siga \sofp \siga
\vphantom{\sp} \right. \nonumber \\
& & \left. \vphantom{\sp}
+ i \siga \g5 \Tr[i \siga \g5  \sofp]
- i \siga \g5 \sofp i \siga \g5 \right] \nonumber \\
& = & \gs \, i \sa \sp \siga \Tr [ \siga \sofp ].
\end{eqnarray}

\noindent Here \Tr denotes the trace over flavour, colour and spinor
indices, and $\sofp$ is the quark propagator
\begin{equalign}
\label{prop}
\sofp^{-1} = \not \! p - \Sigma - m + i\epsilon.
\end{equalign}

\noindent Using the last expression we define the quark condensate
in the following way

\begin{equalign}
\label{cond1}
\langle \qb q \rangle  = -{i \over 2} \Tr S(x=0^-) =
- 4 \nc \, i \sp {\Sigma + m \over p^2 -
(\Sigma + m)^2 + i\epsilon}.
\end{equalign}

\noindent The sum $M = \Sigma + m$ is the constituent
quark mass. Eqs.  (\ref{self1}) and (\ref{cond1}) lead to a
simple relation between $M$ and the condensate 

\begin{equalign}
\label{gap}
M = m - 2 \gs \langle\qb q\rangle.
\end{equalign}

The zeroth-order correlation function is defined by the expression

\begin{equalign}
\label{gcf2}
{\chi}^{(0)}_{AA}(Q) =
2 i \nc \, \Sp \sp [\Gamma_A S(\ppq) \Gamma_A S(\pmq) ].
\end{equalign}

\noindent Here $A=P$ ($A=S$) corresponds to the pseudoscalar (scalar)
channel, $Q^{\mu}=(\omega, {\bf q}\,)$ is the external
momentum, $\Gamma_P=i\g5$, $\Gamma_S=1$, and Sp denotes
the trace over the spinor indices. In the random phase approximation
the full correlation function has the form 

\begin{equalign}
\label{cf1}
{\chi}_{AA}(Q) =
{ {\chi}^{(0)}_{AA}(Q) \over 1 - G_S {\chi}^{(0)}_{AA}(Q) }.
\end{equalign}

\bigskip
It is important to realize that the in-medium correlation function
$\chi_{AA}(Q)$ depends on the variables $\omega^2$  and ${\bf
q}^2 \equiv q^2$ separately. From now on we shall use the notation
$\chi_{AA}(\omega^2,q^2)$ in order to exhibit this dependence
explicitly.  The {\it dynamic mass} is defined by the position of the
lowest lying pole of $\chi_{AA}(\omega^2,0)$. In our case it is
easily found by solving the equation

\begin{equalign}
\label{dm}
1 - G_S {\chi}^{(0)}_{AA}(m_{dyn}^2,0) = 0.
\end{equalign}

\noindent The {\it screening mass}, on the other hand, is defined by the
asymptotic behavior of the static correlation function in r-space,
namely

\begin{equalign}
\label{sm}
m_{scr} = - \lim_{r \rightarrow \infty}
{ d \ln {\chi}_{AA}(r) \over dr},
\end{equalign}

\noindent where

\begin{equalign}
\label{ft}
\chi_{AA}(r) = \int {d^3 q \over (2\pi)^3} \,\, \chi_{AA}(0,q^2) \,\,
e^{i {\bf q} \cdot{\bf r}} =
{1 \over 4 \pi^2 i r} \int\limits_{-\infty}^{+\infty} dq \, q
\,\, \chi_{AA}(0,q^2)\,\, e^{iqr}.
\end{equalign}

\noindent In Eq. (\ref{ft}) the angular integrals were performed using
the fact that $\chi_{AA}(0,q^2)$ is a function of $q^2$ only.  In
vacuum, the dynamic and screening masses are equal due to 
Lorentz invariance. On the other hand, at finite
temperature or density the two masses can be different, since the heat
bath introduces a preferable reference frame. The relation between the
dynamic and the screening masses is discussed more thoroughly in
\cite{FF2}. The main purpose of this paper is to compute the static
correlation functions defined by (\ref{ft}).

\bigskip
\centerline{\bf 3. Physical quantities at finite density}
\bigskip

\bigskip
\noindent{\it i) Imaginary time formalism}
\bigskip

In our calculations we adopt the imaginary time formalism
\cite{FW,JK}. Formally, this is achieved by the following substitution
\cite{JK}

\begin{equalign}
\label{sof}
\int {d^4p \over (2\pi)^4} f(p^0, {\bf p})  \longrightarrow
\int\limits_{-i\infty}^{+i\infty} {dp^0 \over 2\pi}
\int {d^3p \over (2\pi)^3} f(p^0,{\bf p})
+
\oint\limits_{\cal C} {dp^0 \over 2\pi}
\int {d^3p \over (2\pi)^3} f(p^0,{\bf p}).
\end{equalign}

\noindent Here $\cal C$ denotes an integration contour in the complex 
energy plane $p^0$. The position of the contour is fixed by the value
of the chemical potential $\mu$,  see Fig. 1. In the case when
the integrand depends additionally on the external frequency, e.g.,
$f = f(p^0,{\bf p}, \omega)$, the integrals on the right hand side of
formula (\ref{sof}) should be evaluated for purely imaginary values of
$\omega$ and {\it subsequently}  analytically  continued to real
frequencies.

\begin{figure}[ht]
\label{cps}
\xslide{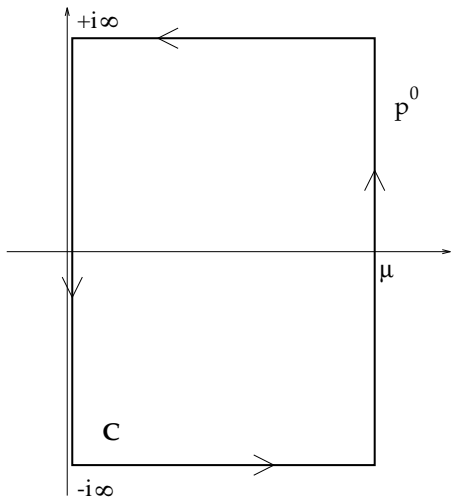}{9cm}{31}{121}{581}{720}
\caption{\small Integration contour in the complex energy plane used for
the evaluation of the matter parts .}
\end{figure}

The advantage of the decomposition (\ref{sof}) is that we can separate the so
called {\it vacuum part} from the {\it matter part}. The vacuum part of a
physical quantity does not {\em explicitly} depend on the occupation of phase
space and reduces at $T = \mu = 0$ to its vacuum expectation value.
On the other hand, the matter part depends explicitly on the occupation of
phase space and consequently vanishes in vacuum. 

Using Eqs. (\ref{cond1}) and (\ref{sof}) we can decompose
the condensate into two parts

\begin{equalign}
\label{cond3}
\langle \qb q \rangle \,\,\, = \,\,\,
\langle\qb q\rangle_{vac} + \langle\qb q\rangle_{mat}.
\end{equalign}

\noindent In the similar way, using Eqs. (\ref{gcf2})  and
(\ref{sof}) we find

\begin{equalign}
\label{psc}
{\chi}^{(0)}_{AA}(\omega^2,q^2) = {\chi}^{(0)}_{AA,vac}(Q^2)
+ {\chi}^{(0)}_{AA,mat}(\omega^2,q^2).
\end{equalign}

\bigskip
\noindent{\it ii) Vacuum parts}
\bigskip

In this subsection we recapitulate our calculation of the vacuum
parts (more details can be found in \cite{FF2}). After a Wick
rotation of the variables $p^0=ip_4$ and $\omega=iq_4$, we find 

\begin{equalign}
\label{condv}
\langle\qb q\rangle_{vac} = -{M \over 2} I_{1, vac}(M^2),
\end{equalign}

\begin{equalign}
\label{ch0pv}
\chi^{(0)}_{PP, vac}(-q_E^2) = I_{1, vac}(M^2)+q_E^2 I_{2, vac}(M^2,q_E^2)
\end{equalign}
and
\begin{equalign}
\label{ch0sv}
\chi^{(0)}_{SS, vac}(-q_E^2) = I_{1, vac}(M^2)
+(q_E^2+4M^2) I_{2, vac}(M^2,q_E^2).
\end{equalign}
\bigskip

\noindent Here $q_E^2\!= q^2+ q_4^2 =
\!q_1^2\!+...+\!q_4^2$ and the functions $I_{1,
vac}(M^2)$ and $I_{2, vac}(M^2,q_E^2)$ are defined by the expressions
below

\begin{equalign}
\label{i1v}
I_{1, vac}(M^2) = 8N_c \int {d^4 p_E \over (2\pi)^4} {1 \over p_E^2 + M^2},
\end{equalign}

\begin{equalign}
\label{i2v}
I_{2, vac}(M^2,q_E^2) = -4N_c \int {d^4 p_E \over (2\pi)^4} {1 \over
[(p_E+q_E/2)^2+M^2][(p_E-q_E/2)^2+M^2]},
\end{equalign}

\noindent where $d^4 p_E = d^3 p dp_4$.

Since the integrals (\ref{i1v}) and (\ref{i2v}) diverge, the
quantities (\ref{condv}), (\ref{ch0pv}) and (\ref{ch0sv}) are not well
defined. In order to obtain finite results, we apply a modified
Pauli-Villars subtraction scheme, where the functions $I_{1, vac}(M^2)$ and
$I_{2, vac}(M^2,q_E^2)$ are replaced by

\begin{equalign}
\label{i1vr}
I_{1, vac}(M^2) \rightarrow I_{1, vac}^R(M^2) = \sum_{i=0}^N A_i I_{1,
vac}(\Lambda_i^2)
\end{equalign}
and
\begin{equalign}
\label{i2vr}
I_{2, vac}(M^2,q_E^2) \rightarrow I_{2, vac}^R(M^2,q_E^2) = \sum_{i=0}^N
A_i I_{2, vac}(\Lambda_i^2,q_E^2).
\end{equalign}

In Eqs. (\ref{i1vr}) and (\ref{i2vr}) $N$ is the number of
subtractions, $A_0=1$ and $\Lambda_0=M$. The coefficients $A_i$, for $i>0$,
have to be chosen in such a way as to provide the finite result for $I_{1,
vac}^R$ and $I_{2, vac}^R$. Moreover, the correlation functions
should be well behaved at infinity (for $q \rightarrow \infty $),
to guarantee the existence of the Fourier transform (\ref{ft}).
These requirements lead to the following set of constraints

\begin{equalign}
\label{ai}
\sum_{i=0}^N A_i = 0, \,\,\,\,
\sum_{i=0}^N A_i \Lambda_i^2 = 0, \,\,\,\, ... \,\,\,\, ,
\sum_{i=0}^N A_i \Lambda_i^{2(N-1)} = 0.
\end{equalign}

Following \cite{FF1,FF2} one finds

\begin{equalign}
\label{ri1v}
I_{1, vac}^R(M^2)={N_c \over 2\pi^2}\sum_{i=0}^N A_i \,
\Lambda_i^2 \, \ln \Lambda_i^2
\end{equalign}

\noindent and

\def\x{{q_E \over 2 \Lambda_i}}
\begin{equalign}
\label{ri2v}
I_{2, vac}^R(M^2,q_E^2)={N_c \over 2 \pi^2}\sum_{i=0}^N A_i
\left[ {2\Lambda_i \over q_E} \sqrt{1+\left(\x\right)^2}
\ln \left(\sqrt{1+\left(\x\right)^2}+\x \right)
+\ln \Lambda_i \right].
\end{equalign}

\noindent Substituting Eqs. (\ref{ri1v}) and (\ref{ri2v}) into
(\ref{condv}), (\ref{ch0pv}) and (\ref{ch0sv}) yields
finite expressions for the vacuum parts of the condensate and the
zeroth-order correlation functions. The correlation functions evaluated
in such a way are
functions of the squared Euclidean momentum. This form is convenient
for doing the Fourier transform (\ref{ft}). In this case
we set $q_4=0$ and we use Eq. (\ref{ri2v}) with the substitution
$q_E^2 = q^2$.

In order to compute the
dynamical masses, we need the function $I^R_{2, vac}(M^2,
-\omega^2)$. This can be obtained by performing the analytic
continuation of the function defined on the right hand side of Eq.
(\ref{ri2v}). The substitution $q_E \rightarrow i\omega \pm \epsilon$
(for $\omega >$ 0) leads to the following result \cite{FF2}

\def\y{\left({\omega \over 2 \Lambda_i }\right)}
\begin{eqnarray}
\label{ri2ac}
&&I_{2, vac}^R(M^2,-\omega^2 \pm i \epsilon) =  {N_c \over 2 \pi^2}
\sum_{i=0}^N A_i \left\{ \Theta(2\Lambda_i - \omega) \left[
{2\Lambda_i \over \omega} \sqrt{1\!-\!\y^2} \hbox{arcsin}\y \right.
\right.
\nonumber \\
&& + \left. \left. \vphantom{\sqrt{\y}} \ln\Lambda_i \right]
 +\Theta(\omega\!-\!2\Lambda_i) \left[{2\Lambda_i \over \omega}
\sqrt{\y^2\!\!\!-\!\!1} \left( \hbox{arcosh}\y \pm {i\pi \over 2}
\right) + \ln\Lambda_i \right] \right\}.
\end{eqnarray}

\bigskip
\noindent{\it iii) Matter parts}
\bigskip

Let us now consider the matter parts of the physical quantities of
interest. In the case of the condensate we find, using Eqs. (\ref{cond1})
 and (\ref{sof}),

\begin{equalign}
\label{condm}
\langle\qb q\rangle_{mat} = -{M \over 2} I_{1, mat}(M^2),
\end{equalign}
where
\begin{equalign}
\label{i1m}
I_{1, mat}(M^2)  =  8i N_c
\oint\limits_{\cal C} {dp^0 \over 2\pi}
\int {d^3 p \over (2\pi)^3} {1 \over p_0^2 -{\bf p}^{\, 2} - M^2}.
\end{equalign}

\noindent To evaluate the function $I_{1, mat}(M^2)$ we collect the
contributions from the poles within the contour $\cal C$ (see Fig. 1)

\begin{equalign}
\label{i1m2}
I_{1, mat}(M^2) = - {2 N_c \over \pi^2} \int\limits_{0}^{\infty}
{dp\, p^2 \over \omega_p} \theta(\mu-\omega_p) =
- {2 N_c \over \pi^2} \int\limits_{0}^{p_F}
{dp\, p^2 \over \omega_p}.
\end{equalign}

\noindent Here $\theta(x)$ is the step function, $\omega_p =
\sqrt{p^2+M^2}$, and $p_F$ is the Fermi momentum of the 
quarks, i.e., $p_F = \sqrt{\mu^2-M^2}$. 

As was pointed out in Section 2,
the matter parts of the zeroth-order correlation functions depend
separately on $\omega^2$ and $q^2$.  Nevertheless, for the extreme
cases of a purely timelike ($Q^{\mu} =
(\omega, 0)$) or a purely spacelike momentum ($Q^{\mu}=(0,{\bf q}\,)$) the
following decompositions, analogous to those obtained in vacuum, are
possible 

\begin{equalign}
\label{ch0pm}
\chi^{(0)}_{PP, mat}(\omega^2,q^2) =
I_{1, mat}(M^2)-Q^2 I_{2, mat}(M^2,\omega^2,q^2)
\end{equalign}
and
\begin{equalign}
\label{ch0sm}
\chi^{(0)}_{SS, mat}(\omega^2,q^2) =
I_{1, mat}(M^2) -(Q^2-4M^2) I_{2, mat}(M^2,\omega^2,q^2).
\end{equalign}
\bigskip

\noindent Here
the function $I_{2, mat}(M^2,\omega^2,q^2)$ is given by the integral

\begin{equalign}
\label{i2m}
I_{2, mat}(M^2,\omega^2,q^2) = 4i N_c \! \oint\limits_{\cal C}
\!{dp^0 \over 2\pi} \! \! \int {d^3 p \over (2\pi)^3}
{ 1 \over \left[(p\!+\!Q/2)^2-M^2\right]
\left[(p\!-\!Q/2)^2-M^2\right]}.
\end{equalign}

\noindent
The calculation of $I_{2, mat}(M^2,\omega^2,q^2)$
proceeds in the same way as the calculation of the function $I_{1,
mat}(M^2)$. Now, as remarked above,
the integral over $p^0$ is evaluated for imaginary $\omega$
and subsequently analytically continued to
real frequencies. The final results (for $\omega >$ 0) are

\begin{equalign}
\label{i2m2}
I_{2, mat}(M^2,\omega^2 \pm i\epsilon,0) &= {N_c \over 2\pi^2}
\int\limits_{0}^{p_F} {dp \, p^2 \over \omega_p} 
{1 \over \omega_p^2-(\omega \pm i\epsilon)^2/4}\\
&= {N_c \over 2\pi^2}\left[\log\left(\frac{p_F+\omega_F}{M}\right)-
{\cal F}(M^2,\omega^2\pm
i\epsilon) \right]
\end{equalign}

\noindent and

\begin{equalign}
\label{i2m3}
I_{2, mat}(M^2,0,q^2) &= -{N_c \over 2q\pi^2} 
\int\limits_{0}^{p_F} {dp \, p \over \omega_p} 
\ln\left|{2p-q \over 2p+q}\right| =\\
&= -{N_c \over 2q\pi^2}{\cal G}(M^2,q^2),
\end{equalign}

\noindent where $\omega_F = \sqrt{p_F^2+M^2}$,

\begin{equalign}
\label{f}
{\cal F}(M^2,\omega^2\pm i\epsilon)
&=-\frac{1}{\omega}\left[\sqrt{4 M^2 - \omega^2} \,\,\arctan
\left(\frac{p_F\omega}{\omega_F \sqrt{4 M^2 -
\omega^2}}\right)\right]\,\,\,\,\,\,\,& (\omega < 2 M)\\
&=-\frac{1}{\omega}\sqrt{\omega^2-4 M^2}\,\,
\ln\left[\frac{\omega_F\sqrt{\omega^2-4 M^2}+p_F\omega}
{\omega_F\sqrt{\omega^2-4 M^2}-p_F\omega}\right]& (\omega > 2 M)\\
& \pm \,i\, {\pi \over 2 \omega}
\theta\left(\omega-2M\right) \theta\left(2\omega_F-\omega\right)
 \sqrt{\omega^2-4 M^2}
\end{equalign}

\noindent and

\begin{equalign}
\label{g}
{\cal G}(M^2,q^2)&= \left(\frac{\sqrt{q^2 +
4 M^2}}{2}-\omega_F\right)\ln\left|\frac{2 p_F+q}{2 p_F-q}\right|\\
&+\frac{\sqrt{q^2+4 M^2}}{2}\,\ln\left[\frac{2 M^2 + q p_F+\omega_F\sqrt{q^2+4
M^2}}{2 M^2 - q p_F +\omega_F\sqrt{q^2+4 M^2}}\right]\\
&-q\ln\left[\frac{p_F+\omega_F}{M}\right].
\end{equalign}

We need
the analytic structure of the function $I_{2, mat}(M^2,0,q^2)$ in
the whole complex $q=|\bf q|$ plane. To this end we represent $I_{2, mat}
(M^2,0,q^2)$ as a sum of two functions, namely

\begin{equalign}
\label{i2m4}
I_{2, mat}(M^2,0,q^2) = I_{2, mat}^{\,(+)}(q) + I_{2, mat}^{\,(-)}(q)
\end{equalign}
where
\begin{equalign}
\label{i2mpm}
I_{2, mat}^{\,(\pm)}(z) = -{N_c \over 4z\pi^2}
\int\limits_{0}^{p_F}
{dp \, p \over \omega_p}
\ln{\,\,\,2p-z \pm i\epsilon \over -2p-z \pm i \epsilon}.
\end{equalign}

\noindent At the end of the calculations we let the infinitesimal
$\epsilon$ go to zero. 
The functions $I_{2, mat}^{\,(\pm)}(z)$ have logarithmic cuts parallel
to the real axis and
stretching from $-2p_F \pm i \epsilon$ to $+2p_F \pm i \epsilon$.
At the respective cut the imaginary part of 
the function $I_{2,mat}^{\,(\pm)}(z)$ is discontinuous. On the
physical Riemann sheet, this amounts to a change in sign: 
\begin{equalign}
\label{iplus2}
\hbox{Im} I_{2, mat}^{\,(+)}(q_R+i\delta_\pm) = \mp {N_c
\over 4q_R\pi} \left[
\sqrt{p_F^2+M^2} - \sqrt{ {q_R^2 \over 4} + M^2} \right],
\end{equalign}
where $q_R$ is real and $-2p_F < q_R < 2p_F$, $\delta_+ =2 \epsilon$
and $\delta_-=0$. Thus,
for the upper sign one is above the cut of $I_{2, mat}^{\,(+)}$ and
for the lower sign below. Obviously the imaginary part of $I_{2,mat}^{\,(-)}$
above and below its cut is equal to that of $I_{2, mat}^{\,(+)}$ above
and below its cut. This can be summarized by the following equation
\begin{equalign}
\label{iplus1}
\hbox{Im} I_{2, mat}^{\,(-)}(z=q_R - i\delta_\mp) &=&
\hbox{Im} I_{2, mat}^{\,(+)}(z=q_R + i\delta_\pm)\\
\hbox{Im} I_{2, mat}^{\,(-)}(z=q_R - i\delta_\pm) &=&
- \hbox{Im} I_{2, mat}^{\,(+)}(z=q_R + i\delta_\pm)
\end{equalign}
Consequently, the cuts are arranged in such a way that
for $z$ on the real axis, the imaginary
part of $I_{2, mat}^{\,(+)}(z)$ cancels that of
$I_{2, mat}^{\,(-)}(z)$ and 
$I_{2, mat}(M^2,0,q^2)$ is real for real $q$ as it should be. 

Using Eqs. (\ref{i2m3}),(\ref{g}),(\ref{i2m4}) and (\ref{i2mpm}) we 
can find the analytic
continuation of $I_{2, mat}(M^2,0,q^2)$ to purely imaginary values of
$q$. Substituting $q=ik$ we find 
\begin{equalign}
\label{i2m5}
I_{2, mat}(M^2,0,-k^2) &=& -{N_c \over k\pi^2} \int\limits_{0}^{p_F}
{dp \, p \over \omega_p}
\arctan \left({2p \over k} \right).
\end{equalign}
The analytic form of $I_{2, mat}(M^2,0,-k^2)$ is easily obtained from
(\ref{i2m3}) and (\ref{g}) by analytic continuation.

The integrals (\ref{i1m}) and (\ref{i2m}), which define the
functions $I_{1, mat}$ and $I_{2, mat}$ are finite. Consequently, 
we are not forced to regularize them. However, if we consider the 
cutoff to be an intrinsic property of the quark-quark interaction, there
is no reason not to regularize the matter parts in the same way as the
vacuum parts. In such an approach, all fermion loops vanish when
$\mu \rightarrow \infty$ (or $T \rightarrow \infty$) due to a
cancellation between the vacuum and matter parts, as was
discussed in the case of finite
temperatures \cite{FF2}. This is a desirable property, since it
guarantees that the condensate and the zeroth-order correlation
functions are well behaved in the high density/temperature limit. 
Consequently, we replace the matter parts by their regularized
counterparts 

\begin{equalign}
\label{i1mr}
I_{1, mat}(M^2) \rightarrow I_{1, mat}^R(M^2) = \sum_{i=0}^N A_i I_{1,
mat}(\Lambda_i^2)
\end{equalign}
and
\begin{equalign}
\label{i2mr}
I_{2, mat}(M^2,\omega^2,q^2) \rightarrow I_{2, mat}^R(M^2,\omega^2,
q^2) = \sum_{i=0}^N A_i I_{2, mat}(\Lambda_i^2,\omega^2,q^2).
\end{equalign}

\noindent
We note that for
$\mu < 2\Lambda_1$, which usually includes the region of interest
to us, the additional terms appearing for $i >$ 0 in Eqs. (\ref{i1mr})
and (\ref{i2mr}) vanish. Therefore, the regularization of the matter
parts become effective only at very large densities. 

The structure of Eqs. (\ref{condv}) -
(\ref{ch0sv}), (\ref{condm}), (\ref{ch0pm}) and (\ref{ch0sm}) suggests
that the following definitions will be useful

\begin{equalign}
\label{i1}
I_1(M^2) = I_{1, vac}^R(M^2) + I_{1, mat}^R(M^2)
\end{equalign}
and
\begin{equalign}
\label{i2}
I_2(M^2,\omega^2,q^2) = I_{2, vac}^R(M^2,q^2-\omega^2)
+ I_{2, mat}^R(M^2,\omega^2,q^2).
\end{equalign}

\noindent Eqs. (\ref{i2m2}) - (\ref{iplus1}), (\ref{i2m5}) have a very
similar form to the expressions found in \cite{FF2}. The formal
correspondence can be obtained by the replacement of the functions
$[\exp(\beta \omega_p) + 1]^{-1}$ by ${1 \over 2}
\theta(\mu-\omega_p)$. Nevertheless, there is a crucial difference
between the present calculation and \cite{FF2}: here the cut along the
real axis is finite, which leads to a completely different behaviour
of the correlation functions at large distances.

\bigskip
\centerline{\bf 4. Density dependence of the  quark and meson masses.}
\bigskip

Using  Eqs. (\ref{gap}), (\ref{cond3}), (\ref{condv}), (\ref{condm})
and (\ref{i1}) we find that

\begin{equalign}
\label{gap1}
M = m + M G_S I_1(M^2).
\end{equalign}

\noindent Eq. (\ref{gap1}) is the gap equation determining the
constituent quark mass $M$. This equation determines
$M$ as a function of the chemical potential $\mu$.
The dynamic masses of pion and sigma are obtained from Eq.
(\ref{dm}). Assuming that the gap equation has a non-trivial solution,
$M \not = 0$, we obtain

\begin{equalign}
\label{pidm}
{m \over M} + m^2_{dyn,\pi} I_2(M^2,m^2_{dyn,\pi},0) = 0
\end{equalign}
and
\begin{equalign}
\label{sidm}
{m \over M} + (m^2_{dyn,\sigma}-4M^2) I_2(M^2,m^2_{dyn,\sigma},0) = 0.
\end{equalign}

\noindent If the current quark mass is zero, the pion is massless, i.e., a
Goldstone boson and the mass of sigma is simply $2M$.

We note that Eqs. (\ref{pidm}) and (\ref{sidm}) are correct not only in
vacuum but also at finite density, as long as a non-trivial solution
to the gap equation exists. Nevertheless, as follows from Eqs. (\ref{ri2ac})
and (\ref{i2m2}),  the correlation functions have cuts for arguments
larger than $2M$. At high densities the meson poles merge with the
quark-antiquark cuts, which means that in this model the mesons can decay into
$q\bar{q}$ pairs. Thus, at these densities there are 
no isolated poles which can be identified with the mass of
the pion or the sigma. We circumvent this difficulty by defining 
the mass as the zero of the real part of Eq. (\ref{dm}), i.e.,

\begin{equalign}
\label{dmm}
1 - G_S \, \hbox{Re} \, {\chi}^{(0)}_{AA}(m_{dyn}^2,0) = 0.
\end{equalign}

\noindent We note that by using Eq. (\ref{dmm}) we implicitly neglect
the $q\bar{q}$ widths of the mesons. This seems reasonable, since
these widths are non-zero only because the Nambu--Jona-Lasinio model
lacks confinement.

\begin{figure}[ht]
\label{qps}
\xslide{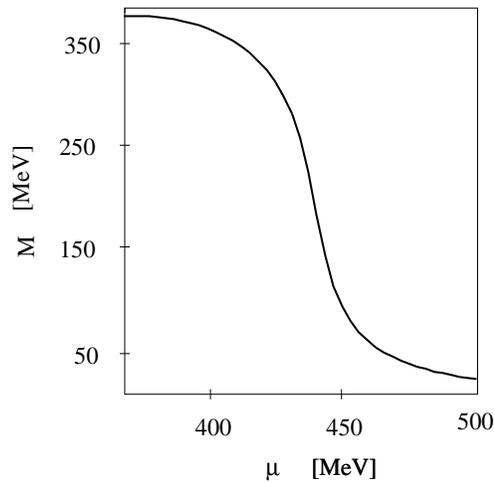}{9cm}{31}{121}{581}{720}
\caption{\small Constituent quark mass $M$ as a function of the quark 
chemical potential $\mu$.}
\end{figure}

In Figs. 2 and 3 we show the density dependence of the constituent quark 
mass and of the dynamic masses of mesons.
We use the same values of the parameters 
as in \cite{FF2}. Consequently, our results for $\mu = 0$
coincide with the $T = 0$ results of \cite{FF2}.
The values of the three regulating masses are: $\Lambda_1 = 680$ MeV,
$\Lambda_2 = 2.1 \Lambda_1$ and $\Lambda_3 = 2.1 \Lambda_2$.
The coupling constant $G_S = 0.75 \, \hbox{fm}^2$ and the current
quark mass $m = 8.56$ MeV. In vacuum we find
the constituent quark mass $M = M_0 = 376$ MeV, the pion mass $m_{dyn,\pi}
= 138$ MeV, and the sigma mass $m_{dyn,\sigma} = 760$ MeV. The corresponding
value of the quark condensate is $\langle \qb q \rangle^{1/3} =$ -- 214 MeV
and of the pion decay constant $f_{\pi}=$ 94 MeV. The latter can be obtained
from the Gell-Mann -- Oakes -- Renner relation: $f^2_{\pi} m^2_{\pi} =
- 2 m \langle {\overline q} q \rangle$.

We note that at $T=0$ all values of the chemical potential in the
range $0 < \mu < M_0$ are equivalent and correspond to the physical
vacuum. The reason is that the chemical potential $\mu$ is the energy
needed to add a quark to the system. Clearly $\mu$ must exceed the
minimal energy, the vacuum rest mass of a quark $M_0$, before the
Fermi sea of quarks begins to be populated.

In Fig. 2, one can see that the constituent quark mass decreases
with the increasing value of the chemical potential. At the same time,
see Fig. 3, the dynamic masses of pion and sigma become
degenerate. This fact indicates restoration of the chiral symmetry at
high density. The overall $\mu$-dependence shown in Figs. 2 and 3
is analogous to the $T$-dependence studied in \cite{FF2}.

\begin{figure}[hb]
\label{dmps}
\xslide{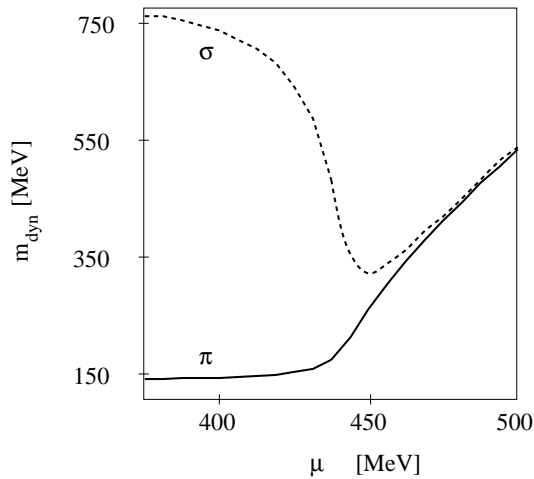}{9cm}{31}{121}{581}{720}
\caption{\small Pion (solid line) and sigma (dashed line) dynamic 
masses plotted as functions of the quark chemical potential $\mu$. }
\end{figure}

The change of the in-medium constituent quark mass $M$ implies, through
Eq. (\ref{gap}), a modification of the value of the quark condensate.
The calculation of the decrease of the condensate with increasing baryon
density in the framework of the NJL model is instructive, since there are
model independent estimates of this quantity, which impose constraints
on the in-medium behaviour of the condensate. Using the
Hellmann-Feynman theorem and the  Gell-Mann -- Oakes -- Renner relation,
one finds \cite{DL,CFG} to leading order in the density

\begin{equalign}
\label{HF}
{ \langle {\overline q} q \rangle  \over
  \langle {\overline q} q \rangle_0 } =
1 -{ \Sigma_{\pi N} \over m^2_{\pi} f^2_{\pi} } \, \rho,
\end{equalign}

\noindent where $\Sigma_{\pi N}$ is the pion nucleon sigma commutator,
$\langle {\overline q} q \rangle_0$ is the vacuum value of the quark
condensate, and $\rho$ is the baryon density. The baryon density
is for two flavours given by $\rho = 2p_F^3/3\pi^2$, where 
$p_F = \sqrt{\mu^2-M^2}$ is the quark Fermi momentum.
We compute the pion nucleon sigma term by using
the relation \cite{CFG}

\begin{equalign}
\label{sigma}
{1 \over 3} \Sigma_{\pi N} =  \Sigma_{\pi q} = m {dM_0 \over m},
\end{equalign}

\noindent where we assume that the $\pi N$ sigma term is simply the
sum of the $\pi q$ sigma terms, like in the
naive quark model.  For the parameters obtained above we find
$\Sigma_{\pi N} = $ 18 MeV, which, using Eq. (\ref{HF}), implies that
the condensate is reduced by 15 \% 
at the saturation density of nuclear matter, $\rho_0 = $ 0.17 fm$^{-3}$.
In Fig. 4  we show
the numerical results for the quark condensate (normalized
to its vacuum value) as a function of the baryon density (normalized to
the saturation density). We note that the calculation agrees with
the low-density theorem (\ref{HF}), when we use the sigma commutator
obtained within the model. Since the sigma term is
much smaller than the empirical value of 45 MeV, the density
dependence of the quark condensate is too weak. Nevertheless, the fact
that Eq. (\ref{HF}) is satisfied, shows that the calculation
is consistent.

\begin{figure}[h]
\label{conps}
\xslide{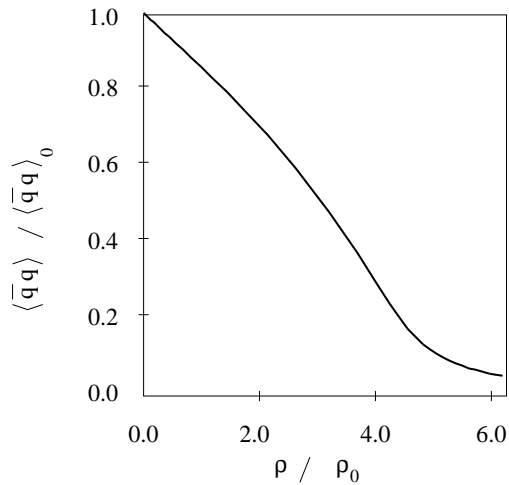}{9cm}{31}{121}{581}{720}
\caption{\small Ratio of the in-medium quark condensate to the vacuum
one, plotted as a function of baryon density.}
\end{figure}

\newpage
\bigskip
\centerline{\bf 5. Static correlation functions in space}
\bigskip
\noindent{\it i) Analytic structure in complex momentum space}
\bigskip

In this Section we discuss our results on the
spatial dependence of static correlation functions. By deforming the
integration contour, the Fourier transform
(\ref{ft}) can be rewritten as a sum of a few contributions,
which are due to the singularities of the
correlation function in the complex $q$-plane. The results of
Section 3 indicate that $\chi^{\,(0)}_{vac}(-q^2)$ has cuts
(in the following we refer to these as the {\it vacuum cuts}) along the
imaginary axis starting at $q=\pm 2iM$ and going to $\pm i \infty$. On
the other hand the function $\chi^{\,(0)}_{mat}(0,q^2)$ has cuts 
({\it matter cuts}) which are parallel to the real
axis. Thus, the full correlation $\chi_{AA}(-q^2)$ function
defined by Eq.
(\ref{cf1}) has at least  these two cuts. Moreover, it can have poles for
imaginary arguments in between the cuts. The general structure of
these singularities is shown in Fig. 5.

\begin{figure}[h]
\label{cutsps}
\xslide{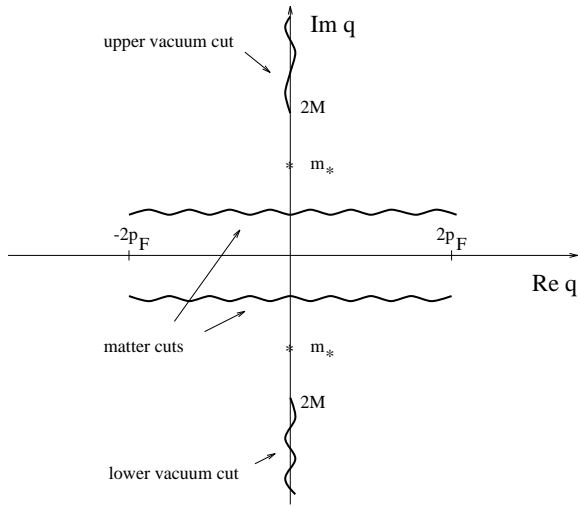}{9cm}{31}{121}{581}{720}
\caption{\small Structure of the singularities of the correlation
functions in the complex $q = |\bf q|$ space. }
\end{figure}

The position of the pole is determined by 

\begin{equalign}
\label{pcontr}
1 - G_S  \chi_{AA}^{(0)}(0,-m^2_{*}) = 0.
\end{equalign}

\noindent A pole can appear only in the interval $0 < m_{\ast} < 2M$,
since the vacuum cut starts at $m_{\ast} = 2M$. In vacuum, due to
the Lorentz invariance, we find that $m_{\ast}=m_{dyn}$.
On the other hand, for $\mu>M_0$,  $m_{\ast} \not = m_{dyn}$ in general,
because the functions $I_{2,mat}(M^2,k^2,0)$ and
$I_{2,mat}(M^2,0,-k^2)$ are different; compare Eqs. (\ref{i2m2}) and
(\ref{i2m5}). The contribution of the pole to the Fourier transform 
is of course of the form $\sim \exp(-m_{\ast}r)$. However, there are
additional contributions from the vacuum as well as the matter cuts.
The integral around the vacuum cut can be written as follows

\begin{equalign}
\label{vaccut}
\chi^{VC}_{AA}(r) = -{1 \over 4 \pi^2 i r} \int_{2M}^{\infty}
dk \, k \left[\chi_{AA}(0,-k^2+i\epsilon)-\chi_{AA}(0,-k^2-i\epsilon) \right]
e^{-kr}.
\end{equalign}

\noindent For large values of $r$, the exponential factor in
(\ref{vaccut}) cuts down the integrand very quickly as a function of
$k$. Thus, the remaining factors, which are slowly varying, can be
approximated by their value at $k \approx 2M$.  Consequently, for $r
\rightarrow \infty$ the contribution of the vacuum cut is also of the
exponential form, $\sim \exp(-2Mr)$.  The contribution of the matter cut is

\begin{equalign}
\label{thermcut}
\chi^{MC}_{AA}(r) = {1 \over 4 \pi^2 i r} \int\limits_{-2p_F}^{+2p_F}
dq \, q \,
\left[\chi_{AA}(0,[q+i\delta_-]^2)-\chi_{AA}(0,[q+i\delta_+]^2)
\right]e^{iqr},
\end{equalign}

\noindent where $\delta_\pm$ is defined below Eq.~(\ref{iplus2}). 
The imaginary shift of the argument, $i\delta_\pm$, denotes that the
function should be calculated just above/below the matter cut.  We
note that in contrast to the finite temperature calculation
\cite{FF2}, the contribution by the matter cut is in the present case
represented by an integral over the finite range in $q$. This leads to
a qualitatively different behaviour of the Fourier transform.

\bigskip
\noindent{\it ii) Results}
\bigskip

Let us start the presentation of the results by discussing the zero
density case ($0 < \mu < M_0$). This is equivalent to the zero
temperature case ($T=0$) studied in \cite{FF2}.  In vacuum the matter
cut vanishes and we are left with only one or two singularities which
contribute to the Fourier transform. This allows for the simple
analysis of the asymptotic behaviour of the correlation functions.

In the case of the pseudoscalar channel we find an isolated pole and a
cut. Because of the Lorentz invariance, the pole coincides with the
dynamical mass. For large values of $r$ the contribution from the
pole, $\sim \exp(-m_{dyn, \pi}r)$, dominates over that of the vacuum
cut, $\sim \exp(-2M_0r)$. Consequently, the static correlation
function in the pseudoscalar channel exhibits a pure exponential
behaviour, whose decay rate is characterized by the pion screening
mass: $m_{scr, \pi} = m_{dyn, \pi} =$ 138 MeV.

In the case of the scalar channel we do not have an isolated pole and
the only contribution to the Fourier transform is due to the vacuum cut.
The latter also leads to the exponential behaviour. In consequence the
sigma screening mass equals $2M_0$ and is a little bit smaller
than the dynamic one computed using Eq. (\ref{dmm}):
$m_{scr, \sigma}=$ 752 MeV and $m_{dyn, \sigma} =$ 760 MeV.

Our numerical calculations, done in the interval 0.5 fm $< r <$ 6 fm,
confirm the results of the analysis carried out above: the correlation
functions in both pseudoscalar and scalar channel decays exponentially
with distance, although the numerically evaluated sigma screening mass
turns out to be slightly larger than $2M_0$. The discrepancy between
the numerical and analytical results is due to the fact that
numerically we are not able to compute at truly asymptotic distances.

Let us now discuss the finite density calculations.  In the case of
the pseudoscalar channel, we did the numerical calculations in the
energy range $M_0$ = 376 MeV $\leq \mu \leq 415$ MeV. In Fig.~6 we
show the corresponding correlation for $\mu = 376, 400$ and 410 MeV.
In contrast to the exponential decay found in vacuum (solid line), the
correlation function at finite density (dashed and dotted lines)
oscillates, with a power-law decay of the amplitude. Its period
decreases with the increasing value of the chemical potential.  From
the physical point of view, such oscillations are caused by the
existence of a sharp Fermi surface.

\begin{figure}[ht]
\label{resps}
\xslide{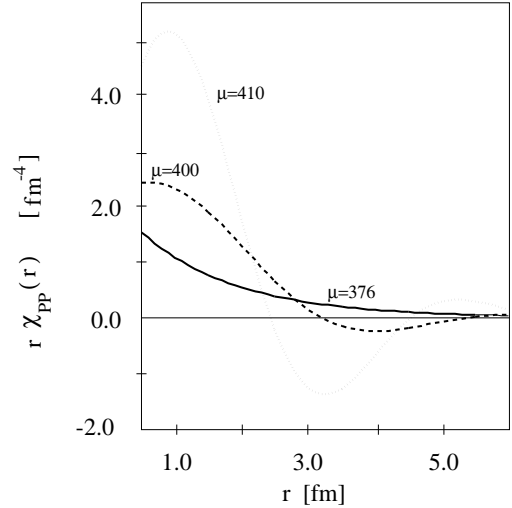}{9cm}{31}{121}{581}{720}
\caption{\small Correlation function in the pseudoscalar channel
plotted for three different values of the chemical potential: $\mu$ =
376 MeV (solid line), $\mu$ = 400 MeV (dashed line), and $\mu$ = 410
MeV (dotted line). }
\end{figure}

The oscillatory behaviour of the correlation functions at finite
density is well known for the non-relativistic degenerate electron
gas, where the phenomenon is called Friedel oscillations
\cite{FW}. These long range oscillations lead to many interesting
phenomena like, e.g, broadening of the nuclear magnetic resonance
lines. A characteristic feature of the Friedel oscillations (at very
large distances) is their period $\delta r = \pi/p_F$. In our case the
situation is similar; for very large values of $r$, the contribution
from the matter cut dominates over the other ones.  Since the cut
extends over a finite range in $q$ ($|q| < 2p_F$) the correlation
function oscillates with the period $\pi / p_F$ (at $r \rightarrow
\infty$).  Nevertheless, the numerical study of the correlation
functions at very large distances is difficult because the
amplitudes decrease with $r$. Consequently, we restrict the numerical
calculations to the interval 0.5 fm $< r <$ 6 fm. 

So far we have discussed the results for $\mu < $ 415 MeV. For larger
values of the chemical potential (corresponding roughly to $\rho >
{1\over 2} \rho_0$) the pseudoscalar correlation function acquires
additional singularity on the real axis for $q\approx 2 p_F$. In the
scalar channel, a similar singularity appears at even smaller
densities ($\rho > {1\over 3} \rho_0$). These singularities indicate
instabilities of the ground state, which lead the system to states of
lower energy.  However, most likely the singularities are artifacts
due to regularization procedure and do not correspond to physical
instabilities. In order to check this point we have redone the
calculation using a different regularization scheme with a 
three-dimensional cutoff. Using the parameters obtained in Ref. \cite{HKZ}, 
we find no singularities on the real axis. This shows that these
singularities are unphysical, since their
presence depends on details in the formulation of
the model. Thus, our regularization scheme can be used only at low
densities. 

On the other hand, the three-dimensional cutoff regularization is
unsatisfactory because it explicitly breaks Lorentz invariance. This
obscures the relation between the correlation functions in timelike
and spacelike regions. Moreover, in the three-dimensional cutoff
scheme, one does not recover the well known screening of the
correlation function at finite temperatures. Instead, the
correlation function oscillates, much like the Friedel oscillations at
$T=0$, with a period $\pi/\Lambda$, where $\Lambda$ is the momentum
cutoff. We stress that this behaviour is an artifact, due to the
finite range of momenta ($|q| < \Lambda$) available in the Fourier
transform. 

Consequently, we stick to the covariant Pauli-Villars
method, in spite of its shortcomings. However, since the results are
inconclusive at high densities, $\rho \sim \rho_0$ and higher, we
restrict the calculations to small densities $\rho < {1 \over
3} \rho_0$. The physics at very large densities will be addressed in
the following section using a different (although closely related) approach.

\bigskip
\centerline{\bf 6.  Perturbative QCD calculation }
\bigskip

The formalism developed so far can also be used to
study correlation functions in perturbative QCD. In this case
the leading term in the meson correlation function is
the lowest order quark loop contribution.
Thus, at large densities the correlation function in
r-space is given by 

\begin{equalign}
\label{htcf}
\chi^{(0)}_{AA}(r) = {1 \over 4 \pi^2 i r} \int\limits_{-\infty}^{\infty}
dq \, q \, \chi^{(0)}_{AA}(0,q^2) e^{iqr}.
\end{equalign}
The Fourier transforms of the vacuum 
$\chi^{(0)}_{AA, vac}(0,q^2)$ and matter $\chi^{(0)}_{AA, mat}(0,q^2)$ parts
define the functions $\chi^{(0)}_{AA, vac}(r)$ and $\chi^{(0)}_{AA, mat}(r)$,
respectively, with

\begin{equalign}
\label{dec}
\chi^{(0)}_{AA}(r) = \chi^{(0)}_{AA, vac}(r) + \chi^{(0)}_{AA, mat}(r).
\end{equalign}

The Fourier transform (\ref{htcf}) can be computed analytically. Using the
methods developed in \cite{FF1}, we find that

\begin{equalign}
\label{htcfvac}
\chi^{(0)}_{PP, vac}(r) = {N_c \over 2 \pi^3 r^3} \sum_{i=0}^N A_i
\Lambda_i^2 \left[3K_2(2\Lambda_i r)+2\Lambda_i r K_1(2\Lambda_i r)\right].
\end{equalign}

\noindent where $K_1$ and $K_2$ are the modified Bessel functions
\cite{AS}. The matter piece, on the other hand, is given by

\begin{equalign}
\label{htcfmat}
\chi^{(0)}_{PP, mat}(r) = {N_c \over 4 \pi^3 r}
\left({2 \over r^3}-{2\over r^2}{\partial \over \partial r} +
{1 \over r}{\partial^2 \over \partial r^2} \right)
\left[G_1(r) + G_2(r) \right],
\end{equalign}
where
\begin{equalign}
\label{g1}
G_1(r) = {\partial \over \partial r} \sum_{i=0}^N {A_i \over 2}
\int_0^{\infty} {dp \over \sqrt{p^2 + \Lambda_i^2}}
\cos(2pr) =
- \sum_{i=0}^N A_i \,\Lambda_i K_1(2\Lambda_i r)
\end{equalign}
and
\begin{equalign}
\label{g2}
G_2(r) = - {\partial \over \partial r} \sum_{i=0}^N {A_i \over 2}
\int_{p_F}^{\infty} {dp \over \sqrt{p^2+\Lambda_i^2}} \cos(2pr).
\end{equalign}

The properties of the modified Bessel functions \cite{GR} imply that
the vacuum part (\ref{htcfvac}) is exactly canceled by the $G_1(r)$
term of the matter part (\ref{htcfmat}).  The cancellation is quite
general and in fact independent of the number of subtractions $N$ and
of the values of the regulating masses $\Lambda_i$.

The vacuum part of the correlation function $\chi^{(0)}_{AA,
vac}(0,q^2)$ is divergent and must be regularized. Nevertheless, the
Fourier transform $\chi^{(0)}_{AA, vac}(r)$ remains finite for $r >
0$ also when one sends the cutoff masses to infinity. To see how this
works we consider the Fourier integral of the regularized vacuum
correlation function. This is well defined, so that the integration
contour can be deformed into a path around the cut on the positive
imaginary axis starting at $p=2 i M$, where $M$ is the quark mass. The
integral around the cut converges for $r > 0$ and the contributions of
the regularization terms, which correspond to cuts starting at $p=2 i
\Lambda_i$ are exponentially suppressed. Consequently, the role of the
regularization terms is to make the Fourier integral finite and thus
to make the deformation of the contour possible. After the contour has
been deformed, the integral is finite, for $r > 0$, even when one
sends the cutoff masses to infinity.
In the limit $\Lambda_i \rightarrow \infty$ and $M
\rightarrow 0$, we find

\begin{equalign}
\label{g2as}
G_2(r) =  {\cos(2p_Fr) \over 2r}
\end{equalign}

\noindent The form of $G_2(r)$ implies that the correlation function
$\chi^{(0)}_{PP}(r)$ is not screened but oscillates in space with a
period $\delta r = \pi / p_F$.  For massless quarks, the behaviour of
$\chi^{(0)}_{SS}(r)$ is identical, since the two channels are
degenerate in this limit.

\bigskip
\centerline{\bf 7. Summary}
\bigskip

In this paper, we have studied the structure of static meson
correlation functions at finite baryon density within the NJL
model. We have restricted our work to the pseudoscalar and
scalar channels. 

In general the vacuum correlation functions in the meson channels are
screened and, due to Lorentz invariance, the screening mass of a
stable meson equals its dynamical mass. For unstable mesons the
screening mass equals the mass of the branch point, since the
asymptotic form of the correlation function picks out the lowest lying
singularity. As demonstrated in our previous work \cite{FF2}, these
results are qualitatively reproduced by the Nambu--Jona-Lasinio model:
the pion screening mass is equal to its dynamic mass, whereas the
sigma screening mass is $2M_0$, where $M_0$ is the constituent quark
mass in vacuum. [The latter result is of course an artifact of
the NJL model, which lacks confinement and of the approximation, which
lacks a coupling of the sigma to the two-pion continuum. In a more
realistic treatment the screening mass in the sigma channel should
equal $2 m_\pi$.] At finite temperature and zero density, the
correlation functions are again exponentially damped, although the
screening and dynamic masses differ. However, as we have demonstrated
here, the correlation functions at $T = 0$ and finite density differ
qualitatively from those in vacuum: they exhibit long ranged
oscillations, of the Friedel type, rather than exponential damping.

In order to understand this effect, we carefully explored the analytic
structure of the correlation functions. We found that the appearance
of the oscillations is connected with the existence of a cut of finite
range in the complex momentum plane. This cut is responsible for the
leading contribution to the correlation function at large distances.
The length of the cut, which is proportional to the Fermi momentum of
the constituent quarks, is reflected in the oscillation period at
large distances $\delta r = \pi / p_F$.  Consequently this form of the
correlation function is quite general and is expected in all normal
Fermi liquids. In particular, the existence of the oscillations is
independent of whether the basic fermionic degrees of freedom are
quarks or nucleons.  Therefore, we feel the oscillatory behaviour of
the correlation function at finite density will not change
qualitatively by confinement.

The fact that the correlation function exhibits Friedel type
oscillations indicates that at finite density it is
impossible to obtain the information about timelike excitations 
by studying spacelike correlation functions. Their long
distance behaviour is dominated by low lying particle-hole
excitations, and consequently not connected with the dynamical mass. 

We find, in agreement with the low density theorem \cite{DL,CFG}, that
the quark condensate is reduced as the density is increased and
eventually almost vanishes, implying that the chiral symmetry is
restored at high densities. This, and the fact that the dynamical
masses of chiral partners (here $\pi$ and $\sigma$) become degenerate
with increasing density, seems to indicate that the NJL model offers a
consistent description of the restoration of chiral symmetry at finite
baryon density.  Nevertheless, the spurious singularities we
encountered in the correlation functions, suggest that the model in
the present formulation cannot be naively used for densities $\rho >
{1 \over 3} \rho_0$. Therefore, we complemented our investigation with
a calculation in perturbative QCD, which again indicates the presence
of oscillations also at very high densities.

{\bf Acknowledgment}\\
One of us (W. F.) would like to thank the Theory Group at GSI for very
warm hospitality. This research was supported in part by the Polish
State Committee for Scientific Research under Grant No. 2 P03B 188 09
and by the Stiftung f\"ur Deutsch-Polnische Zusammenarbeit.

\newpage

\end{large}
\end{document}